\documentclass[pdflatex,sn-mathphys-num]{sn-jnl}
\usepackage{graphicx}%
\usepackage{multirow}%
\usepackage{amsmath,amssymb,amsfonts}%
\usepackage{amsthm}%
\usepackage{mathrsfs}%
\usepackage[title]{appendix}%
\usepackage{xcolor}%
\usepackage{textcomp}%
\usepackage{manyfoot}%
\usepackage{booktabs}%
\usepackage{algorithm}%
\usepackage{algorithmicx}%
\usepackage{algpseudocode}%
\usepackage{listings}%
\usepackage{makecell}
\usepackage{array}
\usepackage{mdwmath}
\usepackage{mdwtab}
\usepackage{eqparbox}
\usepackage{url}
\usepackage{graphicx}
\usepackage{float}
\usepackage{amssymb}
\usepackage[numbers,sort&compress]{natbib}

\theoremstyle{thmstyleone}

\theoremstyle{thmstyletwo}

\theoremstyle{thmstylethree}
\newtheorem{definition}{Definition}

\begin{document}

\title[Article Title]{Tunable Asymmetric Delay Attack in Quantum Clock Synchronization}

\author[1,2]{\fnm{Hui} \sur{Han}}

\author[2]{\fnm{Haotian} \sur{Teng}}

\author[3]{\fnm{Hailong} \sur{Xu}}

\author[2,4]{\fnm{Jinquan} \sur{Huang}}

\author[5]{\fnm{Yuanmei} \sur{Xie}}

\author[3]{\fnm{Yichen} \sur{Zhang}}

\author*[2]{\fnm{Bo} \sur{Liu}}\email{liubo08@nudt.edu.cn}

\author*[1]{\fnm{Wanrong} \sur{Yu}}\email{wlyu@nudt.edu.cn}

\author*[1]{\fnm{Baokang} \sur{Zhao}}\email{bkzhao@nudt.edu.cn}

\author[1]{\fnm{Shuhui} \sur{Chen}}

\affil*[1]{\orgdiv{College of Computer Science and Technology}, \orgname{National University of Defense Technology}, \orgaddress{\city{Changsha}, \postcode{410073}, \state{Hunan}, \country{China}}}

\affil*[2]{\orgdiv{College of Advanced Interdisciplinary Studies}, \orgname{National University of Defense Technology}, \orgaddress{\city{Changsha}, \postcode{410073}, \state{Hunan}, \country{China}}}

\affil[3]{\orgdiv{School of Electronic Engineering}, \orgname{Beijing University of Posts and Telecommunications}, \orgaddress{\city{Beijing}, \postcode{100876}, \state{Beijing}, \country{China}}}

\affil[4]{\orgdiv{School of Electronics and Communication Engineering}, \orgname{Shenzhen Campus of Sun Yat-sen University}, \orgaddress{\city{Shenzhen}, \postcode{518107}, \state{Guangdong}, \country{China}}}

\affil[5]{\orgdiv{College of Science}, \orgname{National University of Defense Technology}, \orgaddress{\city{Changsha}, \postcode{410073}, \state{Hunan}, \country{China}}}
%%==================================%%

\abstract{Quantum clock synchronization underpins modern secure communications and critical infrastructure, yet its fundamental dependence on channel reciprocity introduces an exploitable vulnerability to asymmetric delay attacks.  Current attack strategies rely on static delays, limiting their ability to target application-specific stability requirements. Here, we propose a tunable asymmetric delay attack (T-ADA) that dynamically controls delay parameters to induce manipulate synchronization accuracy. Through experimental implementation, we demonstrate how tailored attack trajectories can selectively compromise system stability across different scenarios. This work uncovers key vulnerabilities in synchronization protocols under customizable attacks and provide a foundation for developing secure and resilient quantum clock synchronization systems.}

 
\maketitle
\section{Introduction}
\label{Introduction}

Precise clock synchronization is crucial for modern communication, navigation systems, financial transactions, and scientific infrastructure~\cite{D5,D3,D4,D6,D7}. Quantum clock synchronization (QCS) leverages the tight two-photon timing correlations inherent in time-energy entangled photon pairs to achieve picoseconds synchronization precision~\cite{A50,A51,A19}, enabling demonstrations over increasingly longer distances and with multiple users~\cite{A2,A15,A47,A7,A43}. While its quantum nonlocality offers resistance to many forms of attacks that disrupt the quantum state, the security of QCS relies on the assumption of reciprocal photon travel times, rendering it vulnerable to asymmetric delay attacks~\cite{A4,A30,A58}. These attacks deliberately manipulate bidirectional transmission times, which often exploiting Faraday rotation within optical circulators to break the channel reciprocity~\cite{A41,A42,A59}.This manipulation degrades synchronization accuracy and compromises system reliability, crucially without disrupting data integrity, making detection inherently difficult.

The disruptive impact of asymmetric delay attacks is not uniform, and critically depends on the specific stability requirements of the target application. For instance, quantum-enhanced telescopes demand exquisite short-term stability~\cite{D13}, rendering them highly sensitive to abrupt timing fluctuations. In contrast, Positioning, Navigation, and Timing (PNT) systems prioritize robust long-term stability~\cite{D12}, making them more vulnerable to slow, accumulating errors. While the work of Lee \textit{et al.} and others confirms the core vulnerability of QCS to asymmetric delays and defenses exist against attacks like intercept-resend~\cite{A1,A32}, a significant limitation existing demonstrations primarily rely on imposing predetermined, static delays. However, a sophisticated adversary aiming to maximize disruption or evade detection would likely employ adaptable strategies, capable of dynamically tuning attack parameters over time to specifically target an application's unique stability requirement (short-term vs. long-term).

To address this problem and enable a comprehensive vulnerability assessment, we present a tunable asymmetric delay attack scheme for quantum clock synchronization systems. The T-ADA enables precise manipulation of channel asymmetry through independently control of three physical parameters: perturbation magnitude, attack duration, and delay trajectories. This parameterization allows the generation of distinct attack patterns—sustained jumps, transient spikes, and gradual drifts—specifically designed to target system. 

We experimentally implement the T-ADA scheme with a round-trip QCS system over 10 km of fiber, achieving a baseline time deviation (TDEV) of $1.85$ ps@$512$ s under normal operation. Critically, applying the T-ADA patterns reveals their targeted disruptive effects. Sustained jumps cause irreversible offsets, degrading long-term stability to $32.05$ ps@$512$ s. Transient spikes induce significant anomalies, worsening short-term instability to $24.88$ ps@$10$ s, but this gradually decreases to $2.36$ ps at an averaging time of $400$ s. Meanwhile, gradual drifts lead to stealthy error accumulation over time, resulting in a TDEV of $68.40$ ps@$1000$ s. These results provide the quantitative and concrete assessment of how tailored asymmetric delay attacks exploit application-specific stability vulnerabilities. Our findings underscore the critical need to develop secure quantum clock synchronization solutions that are resilient to such adaptable threats, ensuring the reliability and security of future clock synchronization technologies.

\section{Results}
\label{Results}

In this experiment, we conducted a 10 km round-trip QCS test, monitoring synchronization between Alice and Bob. The round-trip QCS system is selected for its device simplicity, while enabling monitoring of multiple system performance metrics (e.g., clock difference, one-way/round-trip time difference). The 1550.12 nm pump light emitted by the laser enters the first PPLN waveguide for frequency doubling, and the generated 775.06 nm light then enters the second PPLN waveguide for spontaneous parametric down-conversion (SPDC), producing 1550.12 nm entangled photon pairs. Next, the signal and entangled idler photons are extracted from ITU $\mathrm{CH}$35 (centre at $1549.32$ nm) and $\mathrm{CH}$33 (centre at 1550.92 nm) using a dense wavelength division multiplexer (DWDM). First, the idler photons output from $\mathrm{CH}$33 reach a beam splitter $\mathrm{BS}_1$ and are detected by two superconducting nanowire single-photon detectors (SNSPDs) $\mathrm{D}_1$ and $\mathrm{D}_2$ with 80\% efficiency and 110 ps jitter. The arrival time of the idler photons is recorded by the time-to-digital converter ($\mathrm{TDC}_1$). The TDC's jitter is approximately 8 ps, synchronised to a 10 MHz frequency reference supplied by the rubidium atomic clock (RAC).

Then, the signal photons output from CH35 pass through two optical circulators $\mathrm{OC}_1$ and $\mathrm{OC}_2$, as well as a 10 km optical fiber, reaching a beam splitter $\mathrm{BS}_2$. One output port of the $\mathrm{OC}_2$ is connected to Bob's SNSPD $\mathrm{D}_3$ and recorded by $\mathrm{TDC}_2$, both referenced to a common RAC. The other output port of the $\mathrm{BS}_2$ is connected to port 1 of $\mathrm{OC}_2$, forming a loopback structure. Some photons return to Alice's side, where they are detected and recorded by SNSPD $\mathrm{D}_4$ and $\mathrm{TDC}_1$. To compensate for polarization drift caused by environmental noise, four fiber polarization controllers, $\mathrm{FPC_1}$, $\mathrm{FPC_2}$, $\mathrm{FPC_3}$, and $\mathrm{FPC_4}$, are placed before the SNSPDs. The time correlation measurements between the detection times of the entangled photon pairs are used to sample the one-way and round-trip time differences~\cite{A32,A46,A48}. These entangled photon pairs display strong temporal correlations in their detection events, with the travel time from Alice to Bob governed by the second-order correlation function $G^{(2)}(\tau_{AB})$, which peaks at the time difference $\tau_{AB}$. The round-trip time $G^{(2)}(\tau_{ABA})$ exhibits a similar pattern. Therefore, the clock difference between Alice and Bob is calculated as $\triangle t = \tau_{AB}-\tau_{ABA}/2$.

Here, we use the time deviation (TDEV) as a statistical measure of the short- and long-term stability of system synchronization~\cite{A21}. As shown in Fig.~\ref{Fig2_four_TDEV}, the time deviation (TDEV) confirms the effects of these attack patterns. The baseline TDEV without attack converges more quickly experimentally, reaching $1.85$ ps at an averaging time of $512$ s. The jump attack exhibits significant instability over the measured timescale (up to $512$ s averaging time), with a TDEV of $32.05$ ps. Spike-triggered disruptions show shorter recovery cycles in experiments, with TDEV spiking to $24.88$ ps at $10$ s but gradually decreasing to $2.36$ ps at an averaging time of $400$ s. Notably, gradual attacks demonstrate enhanced concealment in experimental. However, they still cause significant stability degradation, as evidenced by a TDEV of $68.40$ ps at $1000$ s. These three attack patterns collectively reveal that QCS systems are vulnerable across multiple dimensions when facing asymmetric delay attacks.

\subsection{Jump Attack}
\label{Jump Attack}
Jump attacks involve the attacker causing a step-like shift in the clock difference, resulting in a permanent deviation that disrupts the system’s long-term ability to decode time information accurately. In this experiment, we conducted six distinct groups with varying asymmetric delay configurations to investigate the impact of jump attacks in the round-trip QCS system. The control group, with a $0$ ps delay, was compared against five attack groups, each incorporating asymmetric delays of $-10$ ps, $-50$ ps, $-100$ ps, $-200$ ps, and $-500$ ps. Each group was evaluated over a $500$-second measurement interval.

Fig.~\ref{Fig3_Jump_attack} illustrates the impact of five jump-type asymmetric delay attacks, ranging from $-10$ ps to $-500$ ps. In the control group, where no delay ($0$ ps) was introduced, environmental disturbances and device jitter caused the clock difference to fluctuate within about $\pm 50$ ps rather than stay fixed. As the injected delay increased from $-10$ ps to $-500$ ps, the QCS system showed clear, quantised jumps in the clock difference. Specifically, a $-10$ ps delay yielded a measured skew of $-7.9$ ps, while a $-50$ ps delay resulted in a skew of $-48.7$ ps. When the delay increased to $-100$ ps, $-200$ ps, and $-500$ ps, the system recorded skews of $-100.6$ ps, $-195.6$ ps, and $-494.6$ ps, respectively. This experimental setup introduces asymmetry into the transmission paths, with a fixed delay ($A_J$) inserted into the Alice-to-Bob channel, and the delay on the Bob-to-Alice path adjusted to $B_J = -A_J$ to ensure the overall round-trip time remains consistent. This manipulation effectively breaks the symmetry between one-way and round-trip transmission times, resulting in persistent residuals in the clock difference. It demonstrates the potential for subtle jump attacks to evade simple threshold-based detection schemes designed to identify larger anomalies.

\subsection{Spike Attack}
\label{Spike Attack}
The spike attack inserts abrupt and unnatural anomalous values at specific time points in clock difference data, aiming to disrupt or manipulate the normal operation of the QCS system. The impact of such attacks on a system can vary depending on the temporal distribution and frequency of the inserted spikes. In this experiment, we conducted five distinct attacks with varying severity levels, corresponding to different values of spike amplitudes. The amplitudes of the spike attacks were set to $-500$ ps, $-400$ ps, $-300$ ps, $-200$ ps, and $-100$ ps, and each attack was performed independently to observe how the QCS system responds under different conditions.

Fig.~\ref{Fig4_Spike_attack} illustrates the variation in clock offset of the QCS system under spike attacks, with $9900$ ps set as the baseline zero point ($0$ ps reference line). All configured spike attacks induced measurable persistent offsets. Under normal operation, the clock offset should remain stable around the baseline. However, the experiment introduced transient asymmetric delays (ranging from $-500$ ps to $-100$ ps) at specific time points, manifesting as five abrupt negative anomalous pulses highlighted in pink area in the Fig.~\ref{Fig4_Spike_attack}. These attacks were sequentially distributed at $330$ s, $662$ s, $1022$ s, $1376$ s, and $1709$ s. The core disruptive mechanism of the spike attacks lies in the significant deviation of the one-way time difference, which increases with the attack amplitude (experimentally measured deviations ranging from $-494$ ps to $-85$ ps), while the round-trip time difference remains constant. These transient spike attacks demonstrate that even minute but precisely injected delays can substantially compromise the short-term stability of the QCS system.

\subsection{Gradual Attack}
\label{Gradual Attack}
The gradual attack represent an insidious threat, as they aim to subtly manipulate synchronization accuracy over extended periods, evading detection mechanisms sensitive only to abrupt changes. Gradual attacks involve slow, continuous changes in the synchronization accuracy of the system, steadily increasing or decreasing over time from the starting point, with multiple modes of change. In this experiment, we examine the effects of progressive attacks on system clock difference at two different rates of change, as shown in Fig.~\ref{Fig5_gradual_tdev}.

The temporal behavior of the QCS system was analyzed under gradual attack conditions, with a control group operating without any attack, where fluctuations were centered around $-9912.8$ ps, as shown by the blue line in Fig.~\ref{Fig5_gradual_tdev} (a). In the gradual attack scenario, the first attack involved a rate of $-2$ ps per $35$ seconds, where only the Alice to Bob one-way parameter is adjusted. After $2100$ seconds, the system remained unchanged. The second attack occurred at a rate of $-4$ ps per $35$ seconds, where both one-way and round-trip parameters were altered, with $N(t) = -M(t)$. After $1750$ seconds, both parameters reversed direction.

In Fig.~\ref{Fig5_gradual_tdev} (b) compare the TDEV under three conditions: no attack, first attack, and second attack. Under normal conditions with no attack, the system’s time deviation remains stable and decreases steadily over time, indicating the inherent stability and robustness of the QCS system in the absence of external disruptions. In contrast, under the gradual attack scenario, the TDEV initially follows a similar declining trend but eventually diverges from the expected behavior as the attack progresses. Specifically, after an average period of approximately 100 seconds, the TDEV begins to increase, reflecting the destabilizing influence of the continuous delay attacks. As the averaging time increases to $1000$ s, the TDEV increased to $10.70$ ps ($68.40$ ps) for the experiment with the first (second) attack. This upward trend in the TDEV highlights the vulnerability of the system to prolonged asymmetric delay attacks, ultimately leading to a loss of synchronization stability.

\section{Discussion}
\label{Discussion}

Our experimental validation of the tunable asymmetric delay attack (T-ADA) scheme demonstrates its capability to generate distinct attack patterns—jump, spike, and gradual—that dynamically target specific stability regimes in quantum clock synchronization systems. These results extend beyond prior demonstrations of static asymmetric delays, which established feasibility but lacked dynamic targeting~\cite{A1}. This work establishes a formalized model for analyzing asymmetric delay threats through parameterized attack amplitude $A$, timing $t_0$, behavior function $f(t)$, enabling quantitative vulnerability assessment across diverse QCS protocols. Crucially, the complex attacks can be decomposed into fundamental temporal modes, providing essential groundwork for realistic threat modeling and security benchmarking.

A critical insight from our work is the distinct disruptive profile of each attack pattern on QCS stability metrics. Jump attacks induce immediate, permanent offsets. While large jumps cause obvious system failures, even subtle jumps introduce offset errors that can evade initial detection within system noise thresholds. Spike attacks generate severe transient deviations, specifically exploiting and disrupting the high short-term precision application demands. Gradual attacks, in contrast, manifest as a slow, stealthy accumulation of clock offsets, introducing errors persistently yet subtly, making timely detection exceptionally difficult until substantial damage accrues.

Crucially, this differential impact analysis reveals the vulnerability boundaries of synchronization systems and pinpoints blind spots in conventional anomaly detection mechanisms. It demonstrates that the success of an asymmetric delay attack in this dynamic adversary scenario hinges less on the absolute magnitude of the induced offset, and more on whether the attack can introduce deviations in the intended mode. This highlights that smaller, well-calibrated shifts often pose a greater risk due to their covert nature. These findings unequivocally confirm asymmetric delays as a fundamental threat to QCS, echoing concerns in classical clock synchronization~\cite{D14,D15}. 
The results show that quantum enhancements alone are insufficient. To build effective QCS defenses, dedicated security measures informed by an understanding of attack dynamics and detection limitations are essential.

Continuous monitoring of threshold-based defenses cannot be used as a stand-alone detection mechanism. Its primary role is as a mitigation technique to prevent attacks from causing excessive clock drift. While limited in scope, we recognize it may still be one of the few practically effective countermeasures available~\cite{D9,D10}. Another potential defense could be implemented through multi-path redundant design~\cite{D8}. Divide time into discrete slices and randomly allocate the synchronous information transmission paths within each slice. Simultaneously, clock difference collected from new paths undergoes real-time cross-validation against a baseline model established from historical time-series data. The proposed attack schemes provide valuable insights into potential vulnerabilities in quantum clock synchronization, highlighting the need for enhanced defense mechanisms to counteract subtle, progressive delays.

\section{Methods}
\label{Methods}

When a man-in-the-middle attacker introduces an asymmetric delay attack module, the photon signal undergoes different delays during transmission, which will destroy the reciprocity of the synchronization channel. We propose a novel T-ADA scheme that incorporating a dynamic control strategy and hardware module, this scheme provides a comprehensive definition and classification of asymmetric delay attacks, enabling it to address a wide range of QCS systems and attack scenarios with greater versatility and practicality.

Fig.~\ref{Fig6_scheme} illustrates the hardware schematic of the T-ADA scheme. The attack module utilizes two optical circulators ($\mathrm{OC}_1$ and $\mathrm{OC}_2$) to introduce asymmetric delays in the entangled photon transmission path between Alice and Bob. The original fiber length $L=L_1+L_2$ is dynamically adjusted using two motorized optical delay lines (MDLs). By modifying the functions $M(t)$ and $N(t)$, the master controller can precisely control the distance between Alice and Bob, effectively altering the optical path extension or compression.
The T-ADA scheme is shown in Fig.~\ref{Fig7_flow}, which mainly consists of four steps.

Step $1$: Prepared hardware deployment. The hardware deployment environment has been set up in advance. The target optical path has been identified, and the attack hardware module, consisting of two OCs and two MDLs, has been inserted into the target fiber link. The OCs are correctly configured according to the transmission direction. The attack controller is already connected to the hardware, ready to generate attack signals by controlling the MDLs. This setup allows the controller to transmit trigger signals and manage the changes in the delay lines, enabling precise manipulation of the optical path for launching the attack.

Step $2$: System configuration. Determine the parameter configuration based on the type of the target QCS system, which could be a two-way QCS, HOM interference-based QCS, or round-trip QCS. Then, the T-ADA scheme models asymmetric delay attacks by establishing bidirectional photon path delays $M(t)$ and $N(t)$. Let $\alpha, \beta$ represent system-dependent coefficients. The tampered clock difference $\delta$ is defined as:

\begin{equation}
    \delta=\triangle t-\frac{\alpha\cdot M(t)+\beta\cdot N(t)}{2},
\end{equation}
where $\triangle t$ is the calculated original clock difference between Alice and Bob in different QCS systems. $\alpha, \beta$ are parameters that depend on the QCS system configurations. The QCS system configuration parameters are as follows: in the two-way QCS scheme, the parameters are $\alpha$ = 1 and $\beta$ = -1. In the HOM interference-based QCS scheme, the parameters are $\alpha$ = -1 and $\beta$ = 1. In the round-trip QCS scheme, the parameters are $\alpha$ = -1 and $\beta$ = 1.

Step 3: Attack patterns. Asymmetric attacks can manifest in various forms, each inducing distinct anomalies and varying degrees of impact on QCS systems. Given the heterogeneity of attack patterns, a systematic classification of these attacks is essential to gain a comprehensive understanding of their behavior and implications. This paper categorizes asymmetric attacks based on their dynamically operating characteristics over time. This classification method accounts for the temporal evolution of attacks, which can help identify subtle patterns that evolve at different rates or exhibit various forms of disruption throughout the attack's duration. For each attack pattern, the behavior of $M(t)$ and $N(t)$ are defined as
\begin{equation}
    M(N)(t)=A\times f_i(t-t_{0,i})\times \mathrm{H} (t-t_{0,i}),
\end{equation}
where $A$ represents the attack amplitude parameter, $f(t)$ is the function that describes the behavior of an attack pattern, $t_{0,i}$ is the start time of the i-th attack, and $\mathrm{H}(t)$ is the Heaviside step function:
\begin{equation}
    \mathrm{H}(t-t_0)=
\begin{cases}
0,t<t_0 \\
1,t\geq t_0 .
\end{cases}
\end{equation}

The classification of asymmetric delay attacks is based on their temporal evolution characteristics and disturbance intensity distribution: fluctuating perturbations challenge system robustness via long-duration random interference; sudden perturbations disrupt the system's transient response through high-amplitude instantaneous offsets; and progressive perturbations rely on low-intensity sustained offsets to accumulate irreversible errors. These three mechanisms form a complete basis for attack strategies, with any complex attack being decomposed into linear or nonlinear combinations of these fundamental modes.

Case $1$: Jump attacks typically begin at a time point, resulting in a level shift of the clock to increase or decrease. This causes the system to be unable to correctly decode the time information thereafter.

\begin{definition}
\textbf{Jump attack.}
The attacker manipulates the bidirectional delay $M(t)$ and $N(t)$ simultaneously in time $t_0$, causing them to abruptly change to a fixed value and maintain that value thereafter.
\end{definition}
\begin{equation}
    \begin{cases}
        M(t)=A_M^J\times\mathrm{H}(t-t_0) \\
        N(t)=A_N^J\times\mathrm{H}(t-t_0), &
    \end{cases}
\end{equation}
$A_M^J\in \mathbb{R},A_N^J\in \mathbb{R}$ are the attack amplitude in the jump attack. In this scenario, the behavior function is a constant, where $f(t-t_0)=1$.

Case $2$: Spike attacks are characterized by abrupt fluctuations in timestamps, typically manifested as a significant increase or decrease at a specific time $t_0$. These attacks often cause the QCS system to fail in performing accurate time decoding at that moment. Spike attacks are usually fast-moving but can have a severe impact, quickly disrupting the system's accuracy.

\begin{definition}
\textbf{Spike attack.}
The attacker briefly manipulates $M(t), N(t)$ during the time interval $[t_0,t_0+\epsilon]$ creating an instantaneous pulse
\end{definition}
\begin{equation}
    \begin{cases}
        M(t)=A_M^S\times [\mathrm{H}(t-t_0)-\mathrm{H}(t-(t_0+\epsilon))] \\
        N(t)=A_N^S\times [\mathrm{H}(t-t_0)-\mathrm{H}(t-(t_0+\epsilon))], & 
    \end{cases}
\end{equation}
$A_M^S\in \mathbb{R},A_N^S\in \mathbb{R}$ represent the attack amplitude, and $\epsilon$ represents the attack time step, and it is usually controlled by the parameter $\epsilon$ to determine the instantaneity of the spike.

Case $3$: Gradual attacks induce a subtle, progressive increase or decrease in clock differences from the attack’s onset at time $t_0$. This attack pattern can manifest in various modes, such as linear, logarithmic, or polynomial changes in fiber length.
    
\begin{definition}
\textbf{Gradual attack.}
The attacker gradually changes $M(t),N(t)$ starting from $t_0$, creating a smooth delay over time. The specific form of $f_M (t),f_N (t)$ can be chosen based on the gradual attack mode, such as linear, logarithmic, exponential, etc.
\end{definition}
\begin{equation}
    \begin{cases}
        M(t)=A_M^G\times f_M(t-t_0)\times\mathrm{H}(t-t_0) \\
        N(t)=A_N^G\times f_N(t-t_0)\times\mathrm{H}(t-t_0), & 
    \end{cases}
\end{equation}
The amplitude parameters $A_M^G$ and $A_N^G$ control the overall scaling of the attack.

Step 4: Tuning attack operation. The operation of the MDL is dynamically adjusted according to the selected attack pattern, changing the delay $M(t)$ and $N(t)$ in real-time. This includes switching the attack mode, modifying the amplitude, or adjusting the triggering frequency. In an asymmetric delay attack, the attacker can adjust delays in both directions simultaneously or only in one direction. The relationship between$ M(t)$ and $N(t)$ may vary during different attack operations. For example, in HOM interference-based or round-trip QCS schemes, with coordination operations $N(t)=n\cdot M(t)$, where $n$ is a proportional coefficient, and n is typically set as -1 to maintain a constant round-trip photon transmission time. In a two-way configuration, both $M(t)$ and $N(t)$ change independently, either of the same or different types. Finally, the attack controller sends a trigger command to activate the hardware module and start the chosen attack pattern.

{\section*{Data availability}
The raw data supporting this study can be obtained from the corresponding author upon request.}

{\section*{Code Availability}
The code supporting this study can be obtained from the corresponding author upon request.}

\section*{Acknowledgements}

We would like to acknowledge the support from: The Science and technology innovation Program of Hunan Province (2023RC3003), the Hunan Provincial Key Research and Development Program (2025QK3011), and the National Natural Science Foundation of China (U22B2005).

\section*{Author contributions}
B.L. and Y.Z. conceived the idea and designed the experiment. H.H., H.T. and H.X. carried out the experimental work. H.H. and J.H. performed the model analysis. H.H. wrote manuscript writing, with critical feedback and revisions provided by B.L., W.Y., B.Z., S.C., and all other authors. 

\section*{Competing interests}
The authors declare no competing interests.

\bibliography{Bibliography}

\begin{figure}[ht]
    \centering
    \includegraphics[width=1\linewidth]{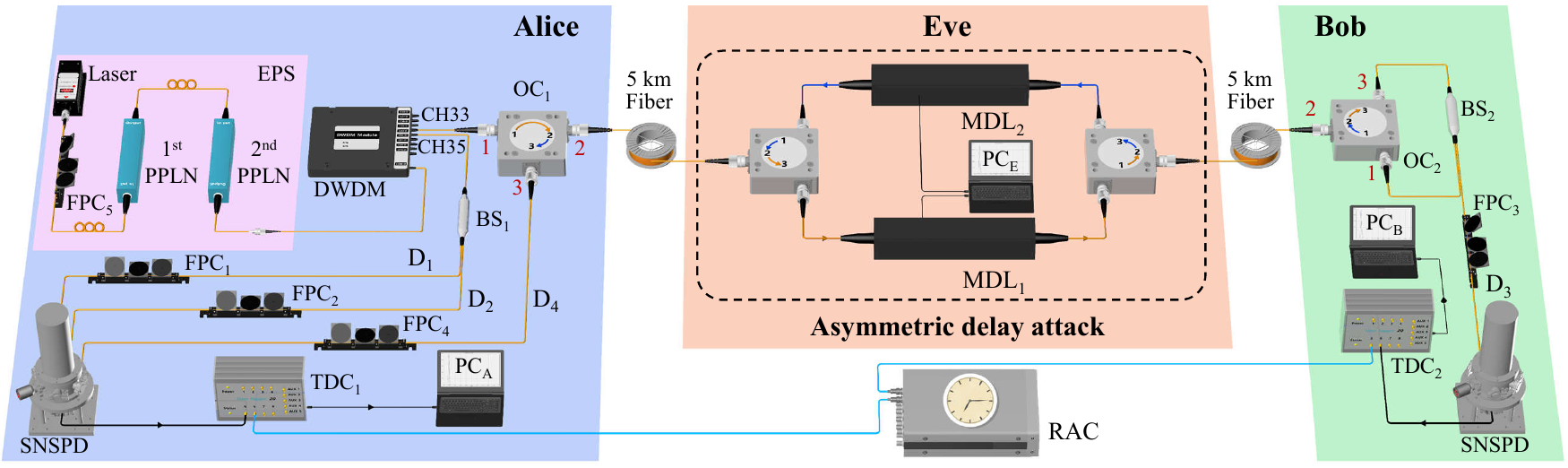}
    \caption{The diagram of the round-trip QCS scheme. FPC: fiber polarization controller, DWDM: dense wavelength division multiplexer, BS: beam splitter, OC: optical circulator, $\mathrm{D_1, D_2, D_3, D_4}$: superconducting nanowire single-photon detector, $\mathrm{TDC}$: time-to-digital converter, $\mathrm{QAC}$: rubidium atomic clock, $\mathrm{MDL}$: motorized optical delay line. $\mathrm{PC}_A$, $\mathrm{PC}_B$ and $\mathrm{PC}_E$ represent Alice's, Bob's, and Eve's personal computers, respectively. $\mathrm{PC}_A$ and $\mathrm{PC}_B$ are statistical photon intensity correlation functions, while $\mathrm{PC}_E$ is configured with a software attack module for the T-ADA scheme.}
    \label{fig:setup}
\end{figure}

\begin{figure}[ht]
    \centering
    \includegraphics[width=0.7\linewidth]{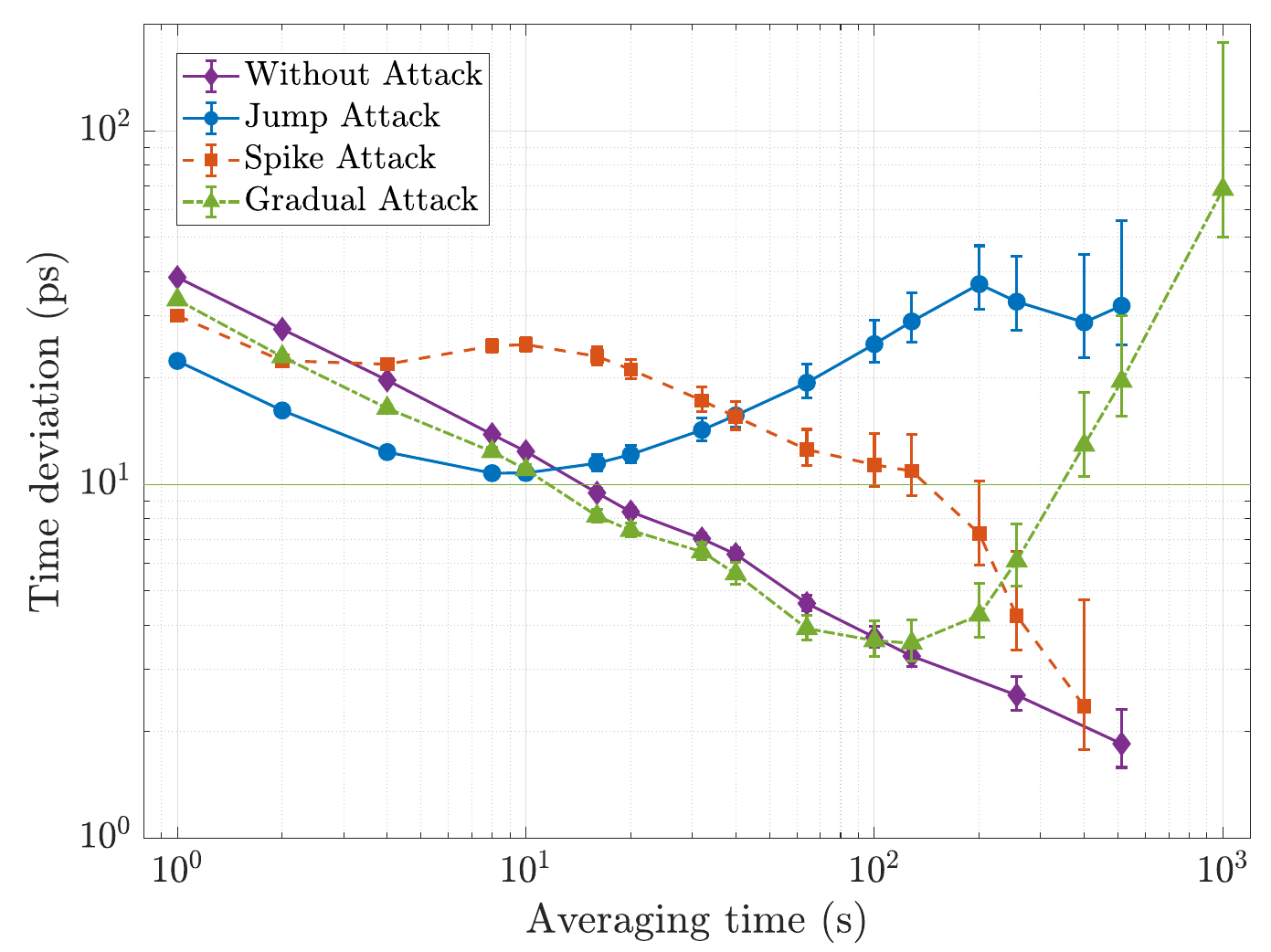}
    \caption{Time deviation under three attack patterns in experimental conditions}
    \label{Fig2_four_TDEV}
\end{figure}

\begin{figure}[ht]
    \centering
    \includegraphics[width=0.75\linewidth]{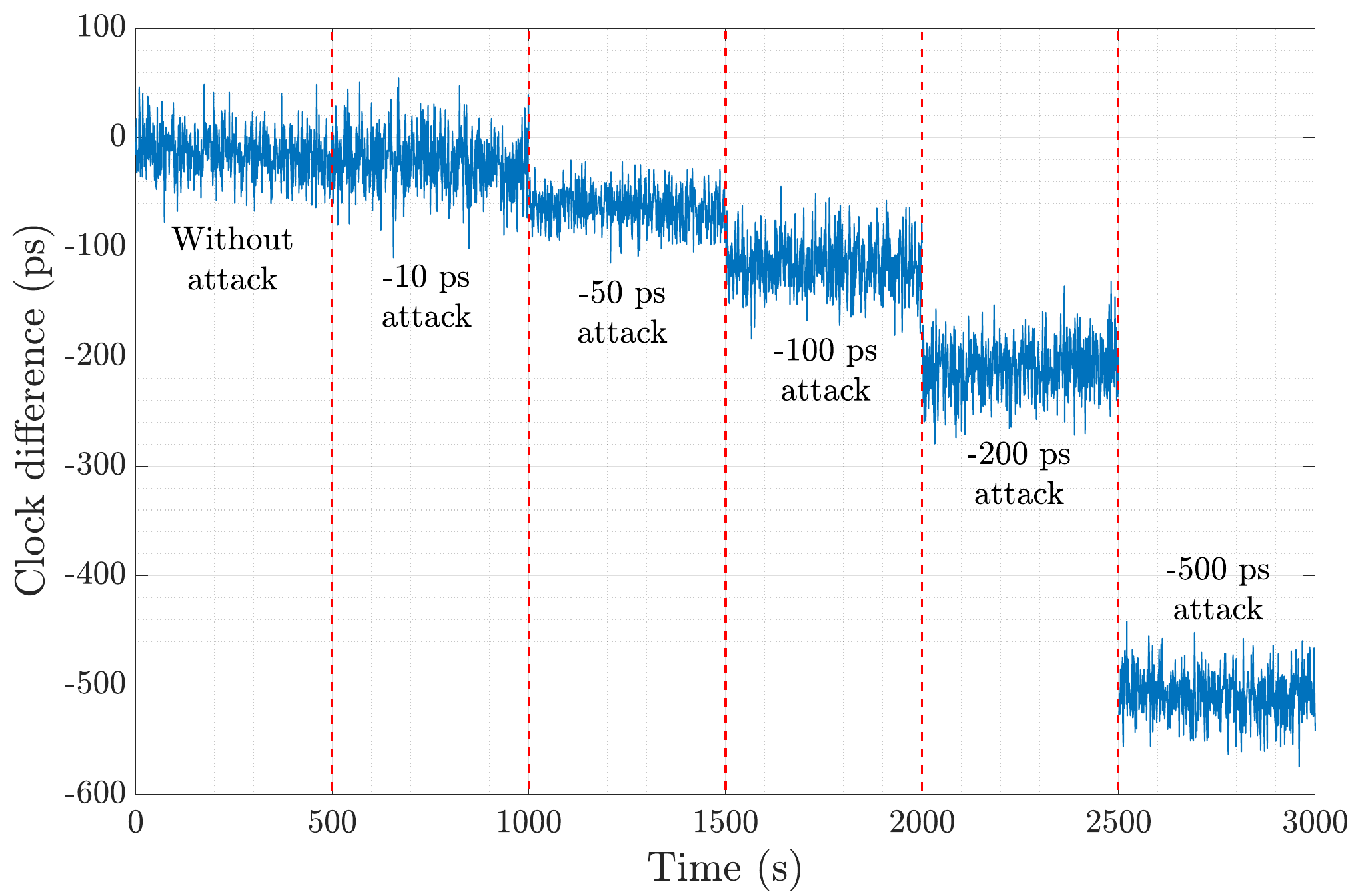}
    \caption{The clock difference of the round-trip QCS system was evaluated across six distinct experimental groups with varying jump attack configurations: $0$ ps, $-10$ ps, $-50$ ps, $-100$ ps, $-200$ ps, $-500$ ps, each measured over a $500$-second time interval. The clock difference was adjusted by adding $-9900$ ps, with this value serving as the reference point ($0$ ps).}
    \label{Fig3_Jump_attack}
\end{figure}

\begin{figure}[ht]
    \centering
    \includegraphics[width=0.75\linewidth]{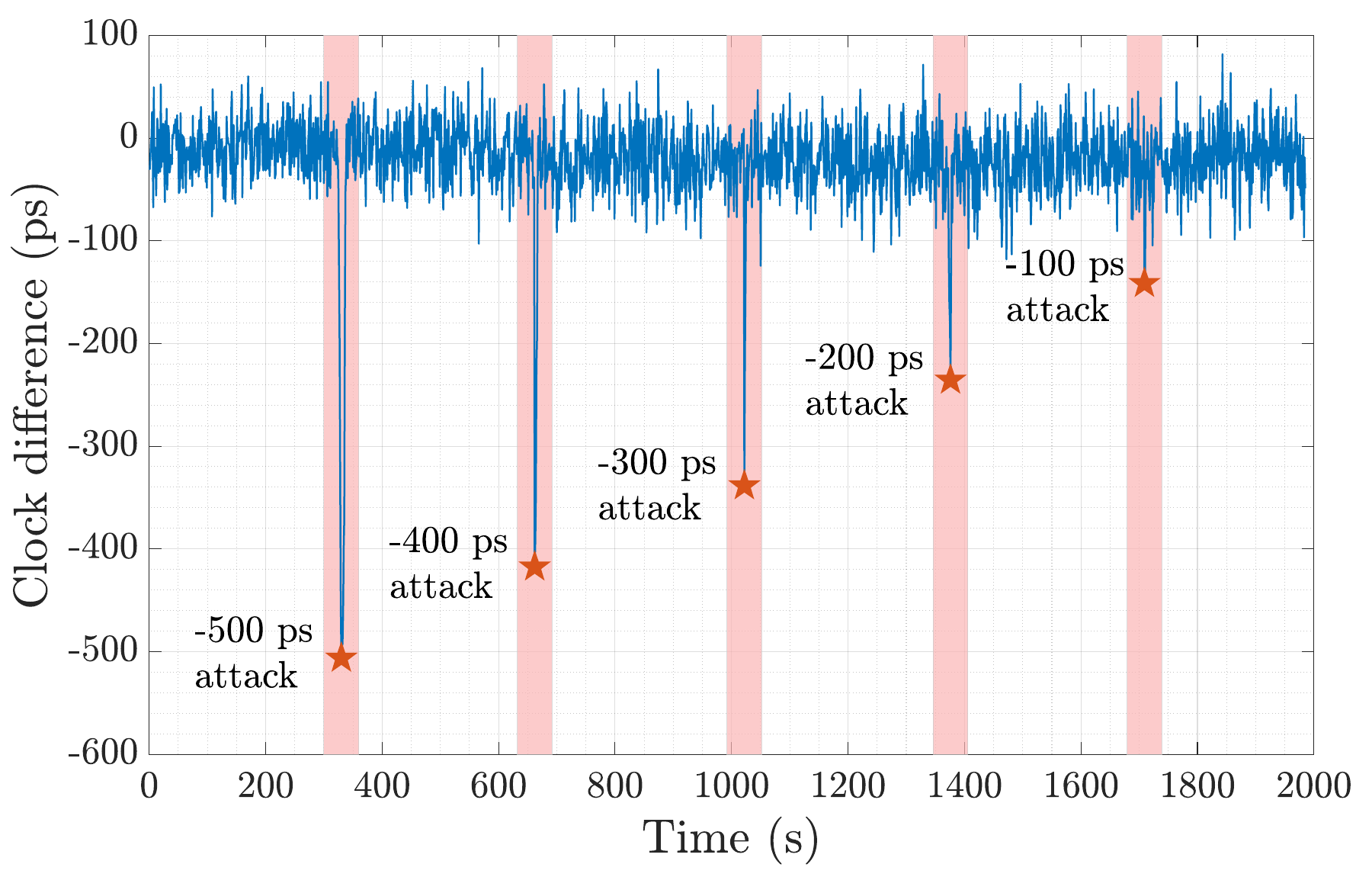}
    \caption{Effects of spike attacks. The clock difference of the round-trip QCS system, with $9900$ ps set as the reference point $0$. The pink region represents the spike attack. The five distinct attacks occurred at different times: $330$ s, $662$ s, $1022$ s, $1376$ s, and $1709$ s, with their corresponding magnitudes increasing from $-500$ ps to $-100$ ps.}
    \label{Fig4_Spike_attack}
\end{figure}

\begin{figure}[ht]
    \centering
    \includegraphics[width=0.75\linewidth]{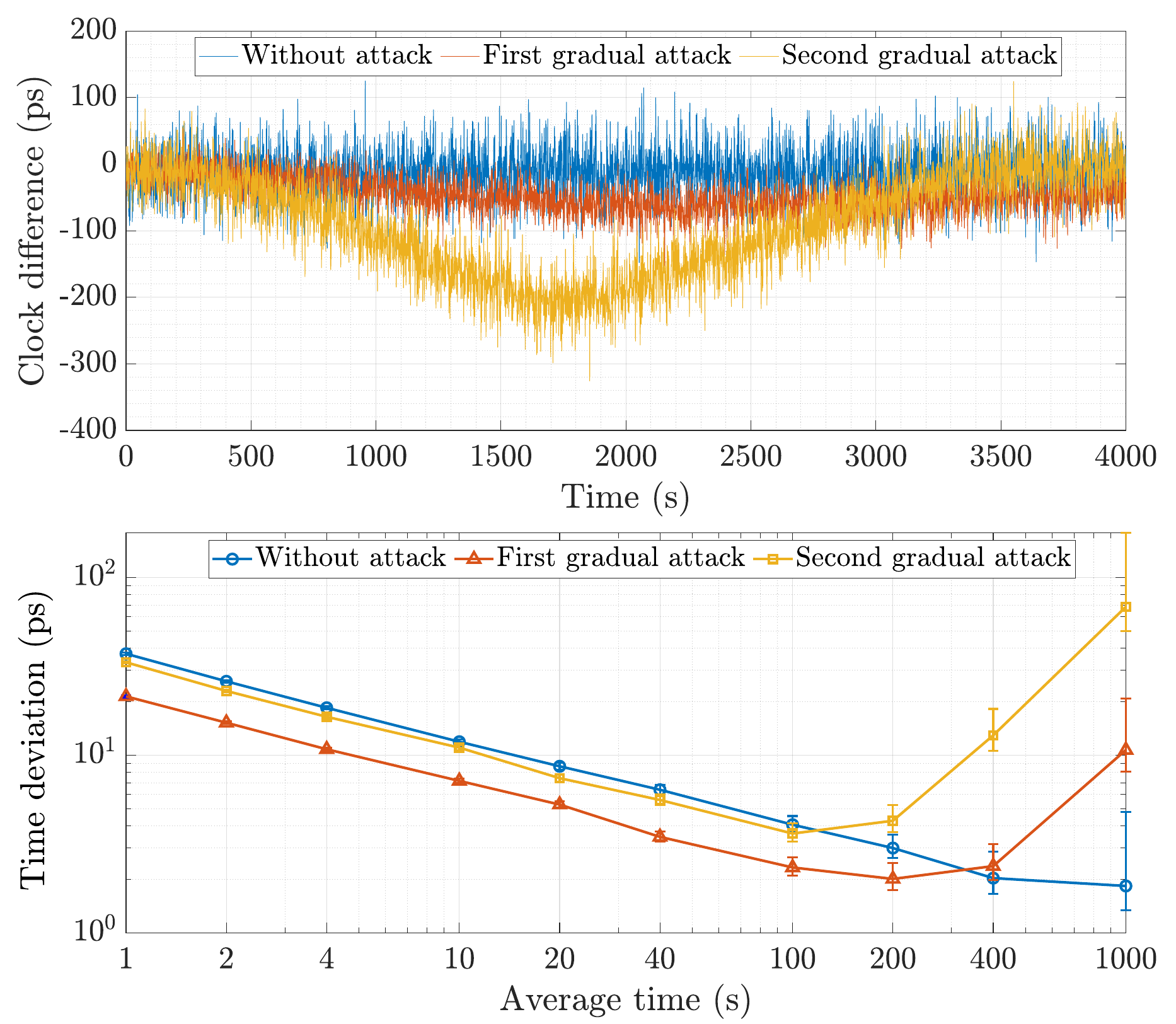}
    \caption{The impact of gradual attacks in experiments, with $9900$ ps set as the reference point $0$. The first attack involves injecting $-2$ ps per $35$ seconds on a unidirectional path, with a total duration of $2100$ seconds and remaining unchanged thereafter. The second attack increases the injection rate to $-4$ ps per $35$ seconds on the unidirectional path, while simultaneously maintaining an opposite injection rate on the round-trip path. After this, the injection rate is adjusted to $4$ ps every $35$ seconds again, extending the attack duration to $3500$ seconds.}
    \label{Fig5_gradual_tdev}
\end{figure}

\begin{figure}[ht]
    \centering \includegraphics[width=1\linewidth]{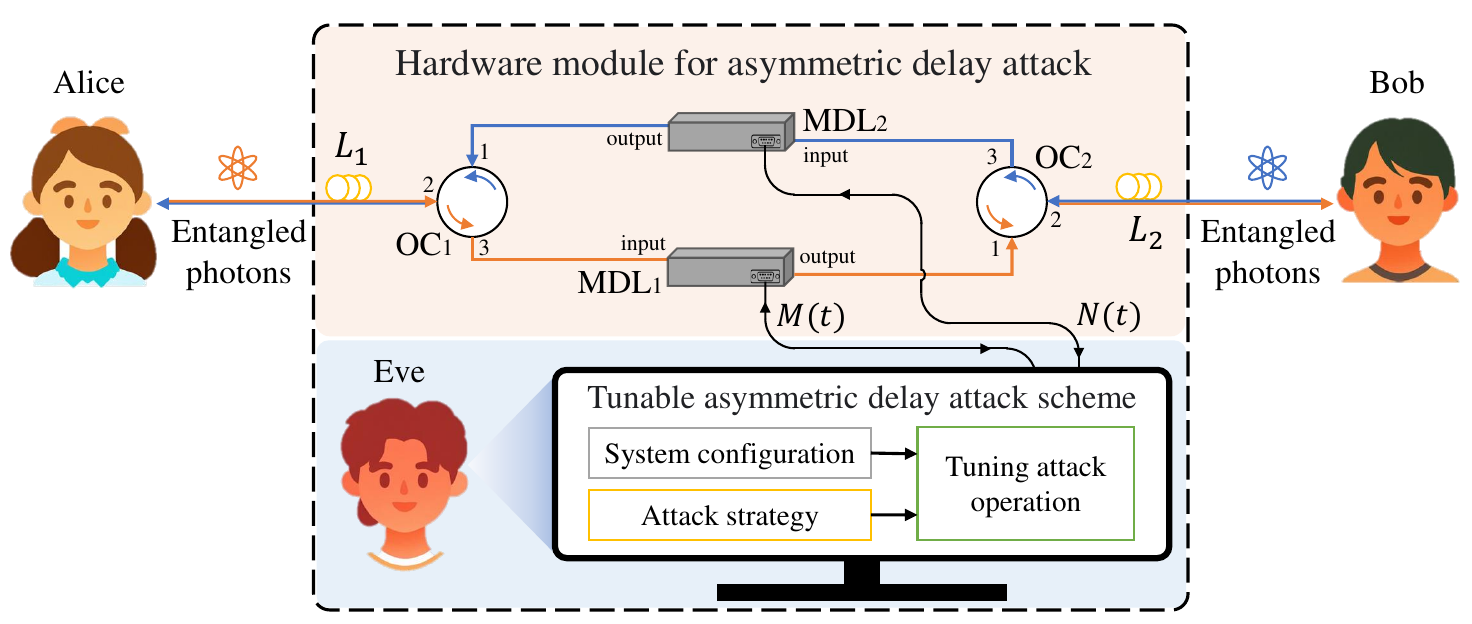}
    \caption{Diagram of the tunable asymmetric delay attack. $\mathrm{OC}$: optical circulator, the number $1$, $2$, $3$ around the $\mathrm{OC}$ represents its port. $\mathrm{MDL}$: motorized optical delay line.}
    \label{Fig6_scheme}
\end{figure}

\begin{figure}[ht]
    \centering \includegraphics[width=1\linewidth]{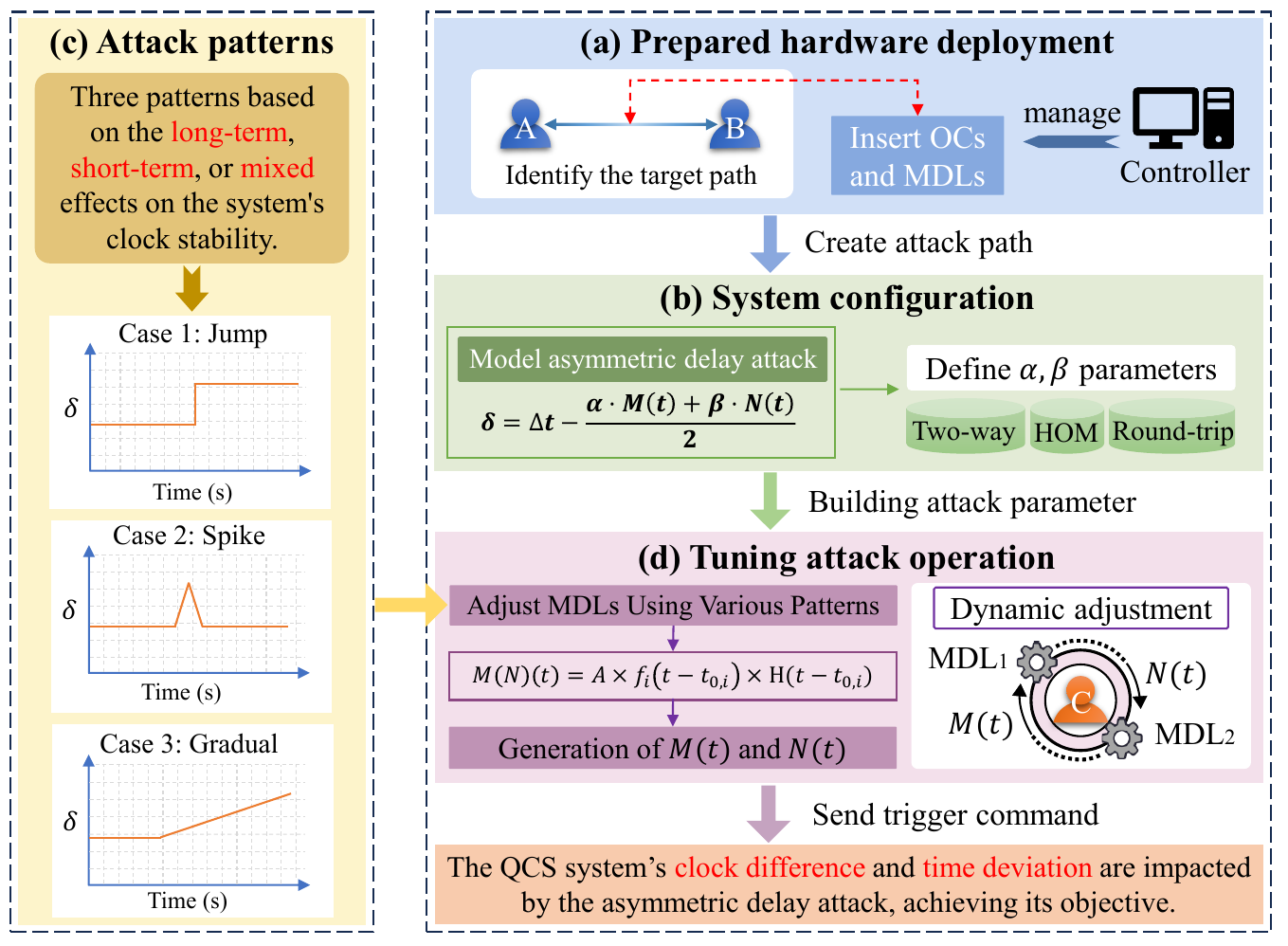}
    \caption{Flowchart of the tunable asymmetric delay attack comprises four sequential phases: hardware deployment, system configuration, attack patterns, and dynamic adjustments.}
    \label{Fig7_flow}
\end{figure}

\end{document}